\documentclass[sigconf,authorversion,nonacm]{acmart}

\usepackage{amsmath}
\usepackage{graphicx} 
\usepackage{amsmath}
\usepackage{adjustbox}
\usepackage{algorithm}
\usepackage{algpseudocode}
\usepackage{enumitem}
\AtBeginDocument{%
  }

\setcopyright{acmlicensed}
\copyrightyear{2018}
\acmYear{2018}
\acmDOI{XXXXXXX.XXXXXXX}

\acmConference[CIKM '24]{}{}
\acmISBN{}

\begin{document}

\title{GPT-Guided Monte Carlo Tree Search for Symbolic Regression in Financial Fraud Detection}

\author{Prashank Kadam}
\email{prashank.kadam@vesta.io}
\affiliation{%
  \institution{Vesta Corporation}
  \streetaddress{5400 Meadows Rd 5th floor}
  \city{Lake Oswego}
  \state{Oregon}
  \country{USA}
  \postcode{97035}
}

\renewcommand{\shortauthors}{Kadam et al.}

\begin{abstract}

  With the increasing number of financial services available online, the rate of financial fraud has also been increasing. The traffic and transaction rates on the internet have increased considerably, leading to a need for fast decision-making. Financial institutions also have stringent regulations that often require transparency and explainability of the decision-making process. However, most state-of-the-art algorithms currently used in the industry are highly parameterized black-box models that rely on complex computations to generate a score. These algorithms are inherently slow and lack the explainability and speed of traditional rule-based learners. This work introduces SR-MCTS (Symbolic Regression MCTS), which utilizes a foundational GPT model to guide the MCTS, significantly enhancing its convergence speed and the quality of the generated expressions which are further extracted to rules. Our experiments show that SR-MCTS can detect fraud more efficiently than widely used methods in the industry while providing substantial insights into the decision-making process.
  
\end{abstract}

\maketitle

\section{Introduction}

Traditionally, financial fraud detection relied on rules constructed by subject matter experts to approve or deny transactions \cite{rule_based}. While these methods were fast and explainable, they struggled to adapt to evolving fraud patterns. Later, machine learning techniques like Logistic Regression \cite{log_reg}, Support Vector Machines \cite{svm}, Random Forest \cite{rf}, and XGBoost \cite{xgb} offered better adaptability but made decision-making less interpretable. Tree-based learners employed algorithms like SHAP \cite{shap} for insights, but these were computationally expensive and challenging to interpret as rules. Advanced Graph Neural Network models like Graph Attention Networks \cite{gat}, Care-GNN \cite{care_gnn}, and Semi-GNN \cite{semi_gnn} improved decision-making but remained black-box and costly. In the recent past, Generative Pre-trained Transformers (GPT) \cite{gpt} have shown path-breaking success in different domains of machine learning. Despite of its success, GPT suffers from issues like hallucinations \cite{hallu_gpt}, limiting their use in critical decision-making.

For real-time applications, fraud detection methods must be fast, interpretable, and accurate. Rule-based methods excel in speed and interpretability. This work introduces SR-MCTS, an algorithm that enhances rule-based methods to match state-of-the-art accuracy. We fine-tune a large language model, Symbolic-GPT \cite{sym_gpt}, with financial data and use it to guide Monte Carlo Tree Search (MCTS) \cite{mcts}. SR-MCTS generates mathematical expressions combining dataset features, operators, and constants to create rule sets for fraud detection. It also minimizes the effect of hallucinations due to limited guidance to the MCTS from Symbolic-GPT.

We evaluate SR-MCTS on our proprietary dataset, showing it outperforms widely used industry methods.

\section{Method}

In this section, we define the Markov Decision Process (MDP) for SR-MCTS, describe Symbolic GPT's role in guiding the search, and outline the reinforcement learning-style approach for evaluating and refining expressions. Finally, we explain how rule sets are extracted from the generated expressions.

\subsection{MDP Formulation}

The symbolic regression problem is modeled as an MDP where:

The \textbf{state space} $S$ consists of all valid combinations of features and constants which form the operands. Operators consist of unary ($\{\sin, \cos, \log, \exp\}$) and binary ($\{+, -, \times, \div\}$) mathematical operations that can be combined with operands to form expressions. The \textbf{action space} $\mathcal{A}_s \subset \mathcal{A}$ for a state $s \in S$ is a conditional set of operators and operands, determined by the current state (e.g., operand followed by operator). The \textbf{transition function} $\mathcal{T}(s, a)$ defines the next state $s'$ given $s$ and action $a$, ensuring mathematically valid expressions. The \textbf{reward function} $R(s, a)$ is inversely proportional to the loss $\mathcal{L}(y, \hat{y})$ of the generated expression, with $y$ as the target and $\hat{y}$ as the prediction.

The goal is to find a sequence of states and actions that maximizes cumulative reward, minimizing the loss function in a reinforcement learning style.

\subsection{PUCT and Symbolic GPT Guidance}

SR-MCTS uses the PUCT (Predictor + Upper Confidence bounds for Trees) \cite{puct} strategy to search the expression space:

\begin{equation}
    \text{PUCT}(s, a) = Q(s, a) + c \cdot \sqrt{\frac{\log N(s)}{N(s, a)}} \cdot \text{P}_{\text{GPT}}(s, a),
\label{eq_puct}
\end{equation}

where $Q(s, a)$ is the expected reward (negative loss) for action $a$ in state $s$, based on previous simulations. $N(s)$ is the number of visits to state $s$, $N(s, a)$ is the number of times action $a$ was selected in $s$, and $\text{P}_{\text{GPT}}(s, a)$ is the probability of selecting $a$ in $s$ as predicted by Symbolic GPT. The constant $c$ balances exploration and exploitation.

\subsection{Evaluation and Loss Calculation}

After generating trajectories (expressions) using SR-MCTS, they are evaluated based on their predictive performance relative to the target variable. For a given expression $\hat{y}$ generated from trajectory $T$, the binary cross-entropy loss is:

\begin{equation}
\mathcal{L}_{\text{BCE}}(y, \hat{y}) = - \left[ y \log(\hat{y}) + (1 - y) \log(1 - \hat{y}) \right]
\label{trx_loss}
\end{equation}

where $y$ is the true value and $\hat{y}$ is the predicted value.

The top $k$ trajectories with the lowest loss are selected, and rewards are assigned as:

\begin{equation}
R(T) = \frac{1}{\mathcal{L}(y, \hat{y}) + \epsilon},
\label{reward}
\end{equation}
where $\epsilon$ avoids division by zero.

\subsection{Symbolic GPT Model and Retraining}

Symbolic GPT, a generative model trained on symbolic regression datasets, is fine-tuned on financial transaction data to guide MCTS. The input to Symbolic GPT is the current state $s$ (a partial expression), and the output is a probability distribution over the next possible actions (operands or operators), effectively serving as the policy $\pi(s)$ for MCTS. This policy guides action selection, enhancing the search process and expression quality.

To further refine Symbolic GPT, the top $k$ trajectories from MCTS are used to retrain the model. The loss function for retraining is cross-entropy with L2 regularization:

\begin{equation}
\mathcal{L}_{\text{CE}} = - \sum_{i=1}^{m} y_i \log(\hat{y}_i) + \lambda \sum_{j} \theta_j^2,
\label{sym_loss}
\end{equation}

where $y_i$ is the true distribution (one-hot encoded) of the next action, $\hat{y}_i$ is the predicted distribution from Symbolic GPT for the $i$-th action, $\theta_j$ are the model weights, and $\lambda$ controls the L2 penalty.

\subsection{Extracting Rules}

After generating the top $k$ expressions using SR-MCTS, we create rules by solving the system of linear equations formed by equating these expressions. By finding the solutions to these equations, we derive interpretable rules that can be used to evaluate transactions as fraudulent or non-fraudulent based on the features present in the dataset.

\begin{algorithm}[h]
\caption{SR-MCTS (Symbolic Regression using MCTS)}
\label{alg:sr_mcts}
\begin{algorithmic}[1]
\Require Pre-trained Symbolic GPT model, Transaction dataset $\mathcal{T}$
\Ensure Optimized set of symbolic regression expressions
\State Initialize an empty set of expressions $\mathcal{T}_{all} \gets \emptyset$
\For{each transaction $t \in \mathcal{T}$}
    \State Initialize an empty set of expressions $\mathcal{T}_t \gets \emptyset$
    \For{each iteration $i$}
        \State Generate an expression $T_i$
        \State Add the expression $T_i$ to  $\mathcal{T}_t \gets \mathcal{T}_t \cup \{T_i\}$
    \EndFor
    \State Evaluate each expression $T_i \in \mathcal{T}_t$ and calculate the loss
    \State Select the top $k$ expressions $\mathcal{T}_{t, \text{top}}$
    \State Update reward $R(T_i)$
    \State Retrain Symbolic GPT with $\mathcal{T}_{t, \text{top}}$
    \State Add the expression $T_t$ to  $\mathcal{T}_{all} \gets \mathcal{T}_{all} \cup \{T_t\}$
\EndFor
\State Repeat the process until convergence
\State Extract rule sets out of the generated expressions
\end{algorithmic}
\end{algorithm}

\section{Experiments}

Our dataset, the Proprietary Financial Fraud Dataset (PFFD), sourced from our organization's e-commerce clients, contains hashed customer order details (e.g., Name, Address, Email, Phone, Device, Payment) to ensure privacy. It covers diverse fraud scenarios, demonstrating the robustness of our approach. Our target label, Fraud Score (fs), ranges from 0 to 100, indicating the likelihood of fraud. The dataset includes approximately 1.25 million transactions, with 13,454 fraudulent cases, making up 1.07\% of the data.

In the experiments, we set $k$ to 0.2, selecting the top 20\% of expressions from each training iteration for fine-tuning Symbolic-GPT. The maximum expression length is 40, with the MDP terminating at the next operand. All categorical variables are one-hot encoded.

We benchmarked SR-MCTS against industry-standard algorithms, including Logistic Regression \cite{hosmer2013applied}, XGBoost \cite{chen2016xgboost}, Random Forest \cite{breiman2001random}, LSTM \cite{hochreiter1997long}, GCN \cite{kipf2016semi}, and GAT \cite{velivckovic2017graph}, training and testing each model on the same datasets.

Results show SR-MCTS surpasses these techniques, while also providing interpretable decision-making.

\begin{table}[htbp]
\centering
\caption{Fraud detection evaluation}
\begin{tabular}{|p{0.15\textwidth}||p{0.05\textwidth}||p{0.05\textwidth}|}
\hline
\textbf{Algorithm} & \textbf{Recall} & \textbf{AUC} \\
\hline
\hline
SVM & 0.536 & 0.507 \\
\hline
Random Forrest & 0.629 & 0.574 \\
\hline
XGBoost & 0.678 & 0.634 \\
\hline
LSTM & 0.621 & 0.587 \\
\hline
GCN & 0.725 & 0.711 \\
\hline
GAT & 0.784 & 0.765 \\
\hline
SR-MCTS & \textbf{0.812} & \textbf{0.797} \\
\hline
\end{tabular}
\label{table1}
\end{table}

\section{Conclusion}

The SR-MCTS approach significantly improves the speed and interpretability of financial fraud detection. Guided by Symbolic GPT, our method achieves faster convergence and generates high-quality expressions that effectively detect fraud. When combined with existing techniques, SR-MCTS enhances performance while maintaining full transparency in decision-making. Future work will refine the model, explore other domains, and experiment with longer expressions, more complex operators, and dynamic detection of $k$ and the terminal state.

\bibliographystyle{ACM-Reference-Format}
\bibliography{sample-base}
\appendix

\section{Neural Networks with MCTS}

Monte Carlo Tree Search (MCTS) \cite{mcts_1} \cite{mcts_2} is widely used for solving combinatorial problems, with its effectiveness enhanced by integrating deep neural networks for value estimation. This approach, known as Neural MCTS, is exemplified in methods like Expert Iteration \cite{ei} and AlphaZero \cite{alpha0}. AlphaZero employs a neural network to approximate both policy and value functions. During learning, multiple self-play rounds are conducted, with MCTS simulations estimating a policy at each state. The selected policy guides the next move, and outcomes are propagated back through the states, with trajectories stored in a replay buffer to train the network.
In each self-play, MCTS runs a fixed number of simulations to generate an empirical policy. These simulations involve four phases:

\begin{enumerate}
\item \textbf{SELECT:} The algorithm begins by selecting a path from the root to a leaf (either a terminal or unvisited state) using an PUCT algorithm \cite{puct}. Starting at the root $s_0$, a sequence of states $\{s_0, s_1, ..., s_l\}$ is determined as follows:

\begin{equation}
    \begin{split}
    & a_{i} = \operatorname{argmax}_a \left[Q(s_{i},a)+c\pi_\theta(s_{i},a)\sqrt{\frac{N(s_i)}{N(s_i,a)}}\right]\\
    & s_{i+1} = \text{move}(s_{i},a_{i})
    \end{split}
    \label{eq:mcts-select}
\end{equation}

Here, $Q$ represents the value of the state-action pair, $N$ is the number of visits to the state-action pair, and $\pi_{\theta}$ is the policy learned by the network. 





MCTS optimizes the output policy to maximize the action value while minimizing deviations from the policy network, assuming accurate value estimates.

\item \textbf{EXPAND:} If the selection phase ends at a previously unvisited state $s_l$, it is fully expanded and marked as visited. All child nodes are considered leaf nodes in the next iteration.

\item \textbf{ROLL-OUT:} Each child of the expanded leaf node conducts a roll-out, where the network estimates the trajectory's outcome. This value is backpropagated to the previous states.

\item \textbf{BACKUP:} The statistics for each node in the selected states $\{s_0, s_1, ..., s_l\}$ are updated.

For the sequence $\{(s_0,a_0),(s_1,a_1),...(s_{l-1},a_{l-1}),(s_l,\_)\}$, the value $V_\theta(s_i)$ for child $s_i$ updates the Q-value iteratively as follows:

\begin{equation}
\begin{split}
    N(s_t,a_t)&\leftarrow N(s_t,a_t)+1\\
    Q(s_t,a_t)&\leftarrow Q(s_t,a_t)+\frac{V_\theta(s_r)-Q(s_t,a_t)}{N(s_t,a_t)}
\end{split}
    \label{eq:mcts-backup}
\end{equation}

This process is repeated for all roll-out outcomes from the previous phase.

\end{enumerate}

After the required number of iterations, the algorithm returns the empirical policy $\hat{\pi}(s)$ for the current state $s$. The next action is sampled from $\hat{\pi}(s)$, and the process continues.

\section{Feature Descriptions}

In this work, we utilize both base and derived features to construct symbolic expressions for financial fraud detection. The features can be categorized as follows:

\subsection{Base Features}
The base features consist of direct transactional attributes that are commonly available during online purchases. Some of these features are:

\begin{itemize}
    \item \textbf{Shipping Features:} 
    \begin{itemize}
        \item \textbf{Shipping Email ($s_{\text{email}}$):} The email address associated with the shipping address of the transaction.
        \item \textbf{Shipping Phone ($s_{\text{phone}}$):} The phone number associated with the shipping address.
        \item \textbf{Shipping Address ($s_{\text{address}}$):} The physical address provided for shipping purposes.
    \end{itemize}
    
    \item \textbf{Billing Features:} 
    \begin{itemize}
        \item \textbf{Billing Email ($b_{\text{email}}$):} The email address associated with the billing address of the transaction.
        \item \textbf{Billing Phone ($b_{\text{phone}}$):} The phone number associated with the billing address.
        \item \textbf{Billing Address ($b_{\text{address}}$):} The physical address provided for billing purposes.
    \end{itemize}
    
    \item \textbf{Other Transactional Features:}
    \begin{itemize}
        \item \textbf{Card Number:} The credit or debit card number used for the transaction.
        \item \textbf{Bin Number:} The bank identification number (BIN) that identifies the institution issuing the card.
        \item \textbf{Device ID:} The unique identifier associated with the device used to complete the transaction.
        \item \textbf{IP Address:} The IP address from which the transaction is conducted.
    \end{itemize}
\end{itemize}

\subsection*{Traditional Velocity Features}
The traditional velocity features capture temporal patterns in transactional activity by calculating the count and sum of base features over different time windows. These windows include: 15 minutes, 30 minutes, 1 hour, 4 hours, 12 hours, 1 day, 7 days, 14 days, 30 days, 60 days, and 90 days. The velocity features help detect unusual activity patterns in a short or long time frame. For example:
\begin{itemize}
    \item \textbf{Count of Shipping Emails ($\text{count}_{s_{\text{email}}}$):} The number of times the shipping email appears in transactions during a specified time window.
    \item \textbf{Sum of Transaction Amounts for a Card Number ($\text{sum}_{\text{bin}}$):} The total sum of transaction amounts associated with the same card number over a specified time window.
\end{itemize}

\subsection*{Relational Velocity Features}
Relational velocity features represent aggregations of one base feature relative to another within a specified time window. This helps capture interactions between different transaction attributes that may indicate fraudulent behavior. For example:
\begin{itemize}
    \item \textbf{Shipping Email vs Billing Address ($\text{rv}(s_{\text{email}}, b_{\text{address}})$):} The number of unique shipping emails associated with a particular billing address in a given time window.
    \item \textbf{Device ID vs Card Number ($\text{rv}(\text{device\_id}, \text{card\_number})$):} The number of times the same device ID is used with different card numbers in a specific time window.
\end{itemize}

These base and velocity features form the core input for constructing the symbolic expressions generated through the SR-MCTS process. By considering both temporal patterns and relationships between features, the model can detect fraudulent activities more effectively and interpretably.

\section{Generated Expressions}

Following is an example of the expression generated by Symbolic MCTS:

\begin{equation}
\begin{aligned}
\text{Fraud Score} = & \ 0.31 \log\left(1 + \text{count}_{s_{\text{email}}}^{30 \ \text{day}}\right) \\
& + 0.54 \log\left(1 + \text{sum}_{b_{\text{address}}}^{15 \ \text{min}} 
+ \text{count}_{\text{card\_number}}^{1 \ \text{hr}}\right) \\
& + 1.21 \sin\left(\text{sum}_{\text{bin\_number}}^{1 \ \text{day}} 
\cdot \text{count}_{\text{device\_id}}^{4 \ \text{hr}}\right) \\
& + \exp\left(-0.77 \left(\text{sum}_{s_{\text{phone}}}^{7 \ \text{day}} 
+ \text{count}_{\text{ip}}^{90 \ \text{day}}\right)\right) \\
& + 0.93 \left(\text{count}_{b_{\text{email}}}^{12 \ \text{hr}} 
+ \text{sum}_{\text{device\_id}}^{30 \ \text{day}}\right) \\
& - 2.53 \log\left(1 + \left(\text{sum}_{s_{\text{email}}}^{60 \ \text{day}} 
+ \text{count}_{b_{\text{address}}}^{90 \ \text{day}}\right)\right) \\
& + 0.26 \left(\text{count}_{s_{\text{phone}}}^{1 \ \text{hr}} 
+ \text{count}_{\text{ip}}^{30 \ \text{day}}\right)^2 \\
& - 1.84 \log\left(1 + \exp\left(\text{sum}_{b_{\text{email}}}^{4 \ \text{hr}} 
- \text{sum}_{s_{\text{address}}}^{1 \ \text{day}}\right)\right) \\
& + 0.41 \log\left(1 + \text{rv}\left(s_{\text{email}}, b_{\text{address}}, 30 \ \text{day}\right)\right) \\
& - 0.68 \exp\left(\log\left(1 + \text{sum}_{\text{card\_number}}^{1 \ \text{day}} 
+ \text{count}_{\text{device\_id}}^{7 \ \text{day}}\right)\right) \\
& + 1.31 \sin\left(\log\left(1 + \text{count}_{s_{\text{phone}}}^{90 \ \text{day}}\right) 
+ \log\left(1 + \text{sum}_{\text{ip}}^{30 \ \text{day}}\right)\right) \\
& - 0.83 \left(\text{count}_{\text{bin\_number}}^{4 \ \text{hr}} 
+ \log\left(1 + \text{rv}\left(\text{dev\_id}, \text{bin}, 
60d \right)\right)\right) \\
& + 0.25 \left(\text{count}_{s_{\text{address}}}^{12 \ \text{hr}} 
+ \log\left(1 + \text{sum}_{b_{\text{email}}}^{30 \ \text{day}}\right)\right)^2 + 0.27 \\
\end{aligned}
\end{equation}

The coefficient of each term has been rounded-off to the nearest second digit after the decimal. ${rv}$ in the above equation stands for relational velocity. A rule can be further created from the above equation assigning the Fraud Score to be 1 and creating inequalities to trigger the rules.

\end{document}